\input epsf 
\documentclass[preprint,eqsecnum,aps]{revtex4}
\usepackage{graphicx}
\def\beq{\begin{eqnarray}}
\def\eeq{\end{eqnarray}}

\begin{document}
\title{High order analysis of nonlinear periodic differential equations}
\author{Paolo Amore~\footnote{paolo@ucol.mx} and H\'ector Montes Lamas}
\affiliation{Facultad de Ciencias, Universidad de Colima, Colima, M\'exico}
\date{October 15, 2003}
\vskip .1in
\thispagestyle{empty}

\begin{abstract}    
In this letter we apply a method recently devised in \cite{aapla03} to find precise approximate solutions 
to a certain class of nonlinear differential equations. The analysis carried out in \cite{aapla03} is refined 
and results of much higher precision are obtained for the problems previously considered (Duffing equation, 
sextic oscillator). Fast convergence to the exact results is observed both for the frequency and for the 
Fourier coefficients. The method is also applied with success to more general polynomial potentials 
(the octic oscillator) and to the Van Der Pol equation.
\end{abstract}                                                

\maketitle

\section{Introduction}
\label{sec:intro}

In this letter we extend the analysis of \cite{aapla03}, where a new method for the solution of 
oscillatory nonlinear problems was devised. This method works by combining the Lindstedt-Poincar\'e (LP)~\cite{Lin83}
method to the  Linear Delta Expansion (LDE)\cite{lde}, or Optimized Perturbation Theory (OPT)~\cite{Yuk91}.
A detailed list of references on the LDE can be found in \cite{aapla03}.

The LP method is used to solve approximately nonlinear differential equations with periodic solutions, by
introducing strained coordinates, which allow to eliminate the secular terms arising in the 
perturbative expansion;  in the LDE method an insoluble theory is interpolated with a soluble one depending 
on some arbitrary parameter: an expansion is then carried out in terms of a suitably defined ``perturbation''.
Since the ``perturbation'' is not a priori fixed and not necessarily expressed in terms of ``small'' 
parameters, the method is truly nonperturbative.
 
As proved in \cite{aapla03} the application of LDE to the LP method, which is at the core of our method, 
allows to extend the analysis to the nonperturbative regime (large nonlinearities) and to obtain errors 
which are small even in regimes where the LP method fails completely. 

In this letter we pursue two goals: on one hand to investigate the convergence of our method
in the cases previously analyzed  to higher orders; on the other hand to test the method  by applying it 
to more demanding problems, such as the Van der Pol equation. 

The letter is organized as follows: in Section~\ref{sec:duff} we  study the Duffing equation and discuss 
the convergence of the method; in Section~\ref{sec:anharm} we analyze more general anharmonic potentials 
(sextic and octic); finally, in Section~\ref{sec:vdp} we consider the Van der Pol equation. Finally in 
Section~\ref{sec:concl} we draw our conclusions.

\section{Duffing equation}
\label{sec:duff}

We start our analysis by considering the Duffing equation, which describes the oscillations of a unit mass in a potential
$V(x) = \frac{x^2}{2} + \frac{\mu \ x^4}{4}$. It reads:
\beq
\frac{d^2x}{dt^2}(t) + x(t) = - \mu \ x^3(t) \ .
\label{duf1}
\eeq
$\mu$ is a coupling which controls the strength of the nonlinear term. 
The application of the LPLDE method requires to write eq.(\ref{duf1}) as 
\beq
\Omega^2 \ \frac{d^2x}{d\tau^2} + \left( 1 + \lambda^2 \right) \  x(\tau) = \delta \ \left[ - \mu \ x^3(\tau) 
+ \lambda^2 \ x(\tau) \right] \ ,
\label{duf2}
\eeq
where $\lambda$ is an arbitrary parameter and $\Omega$ is the exact frequency of the system. We also introduced a power-counting
parameter $\delta$, which allows to recover eq. (\ref{duf1}) when the value $\delta = 1$ is taken. The r.h.s. of eq.(\ref{duf2})
is then treated as a perturbation, although its size is not fixed, being $\lambda$ completely arbitrary. When the perturbative 
expansion is carried out to finite order, a spurious dependence upon this parameter will show up in all the observables. 
The Principle of Minimal Sensitivity (PMS) will then be used to choose a $\lambda$ which minimizes this dependence. This 
procedure is explained in detail in \cite{aapla03}. 

\begin{figure}
\includegraphics[width=9cm]{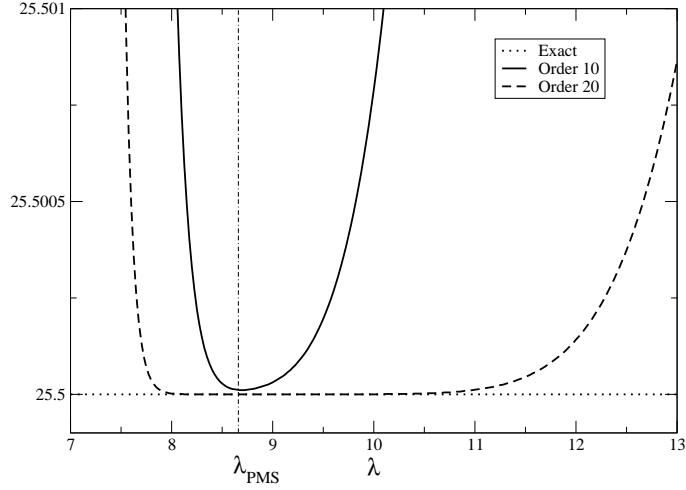}
\caption{Energy of the Duffing oscillator corresponding to $x = 0$.The vertical line is the optimal value to third order. We assume $\mu = 100$
and $A = 1$.}
\label{Fig000}
\end{figure}

To a given order $N_{max}$ we write the frequency and the solution as
\beq
\Omega^2 = \sum_{n=0}^{N_{max}} \ \delta^n \ \alpha_n \ \ \ , \ \ \ x(t) = \sum_{n=0}^{N_{max}}  \ \delta^n \ x_n(\tau)  \ .
\label{omegax}
\eeq
and we take $\delta = 1$ at the end. Both the coefficients $\alpha_n$ (which are obtained by eliminating the resonant contributions at
any given perturbative order) and the solutions $x_n(\tau)$ are $\lambda$-dependent.
The dependence upon the arbitrary parameter is then eliminated by imposing the PMS condition  $\frac{d\Omega^2}{d\lambda}=0$. 
Remarkably we have observed that the solution to this equation to any given order is approximated extremely well by the 
third order solution, i.e. $\lambda = \frac{\sqrt{3 \ \mu} \ A}{2}$, both for positive and negative values of $\mu$. 
This allows us to obtain expressions which are completely {\sl analytical}. We can also attach a more physical meaning 
to the PMS condition by looking at Fig.~\ref{Fig000}: in this Figure we plot the energy of the oscillator evaluated at $x = 0$, 
where all the energy is kinetic, and compare it with the exact expression for the energy, which is given by the potential energy 
at the inversion point. The approximate solution found with eq.~(\ref{omegax}) violates the conservation of energy by an amount which 
depends upon $\lambda$. However, when the optimal value of $\lambda$ is used (the vertical line in the Figure) such violation is nearly minimal.

\begin{figure}
\includegraphics[width=9cm]{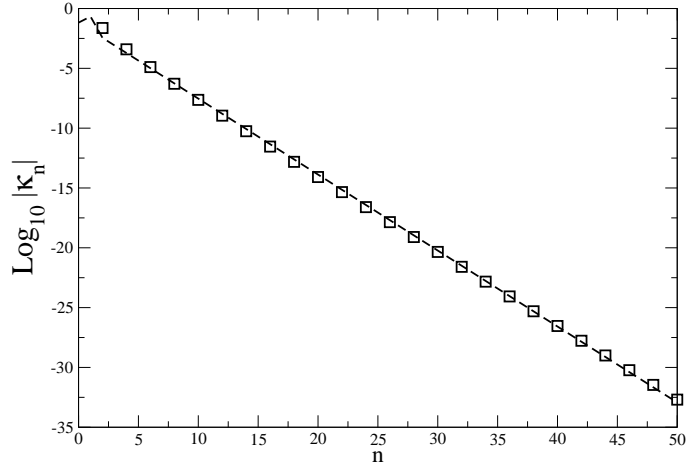}
\caption{Coefficients of the formula (\ref{kappa}). The dotted line is a linear fit.}
\label{Fig0}
\end{figure}

We have noticed that the coefficients $\alpha$ of eq. (\ref{omegax}) can be written as:
\beq
\alpha_0 = 1 + \frac{3}{4} \ A^2 \ \mu \ \ \ , \ \ \ 
\alpha_{2 n} = - \frac{\kappa_{2 n} \ \left( A^2 \ \mu\right)^{2 n}}{\left(1 + \frac{3}{4}  \ A^2 \ \mu \right)^{2 n-1}} \ \ \ , \ \ \ 
\alpha_{2 n+1} = 0
\label{alpha}
\eeq
where the $\kappa_n$ are purely numerical coefficients and $n = 1,2, \dots$. Such relations hold for all the $\alpha$ that we have calculated.
In Fig.~\ref{Fig0} we plot the logarithm of the coefficients, for $N_{max} = 50$; the  dashed line is a linear fit corresponding to:
\beq
\kappa_{n} = 0.0663 \cdot e^{-1.46225  \ n}
\label{kappa}
\eeq
with $n$ even. The first few coefficients are written in Table \ref{table1}.

\begin{figure}
\includegraphics[width=9cm]{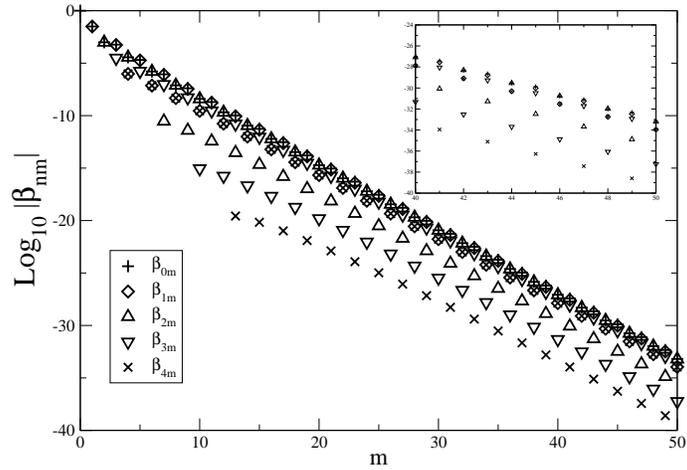}
\caption{Logarithm of $\beta$ corresponding to different Fourier coefficients and to different orders.}
\label{Fig00}
\end{figure}

The expression in eq. (\ref{omegax}) can be written equivalently as
\beq
x_{approx}(t) &=& \sum_{n=0}^{N_{max}} \ c_n^{(approx)} \ \cos \left[(2 n+1) \ \Omega \ t\right]  \ .
\label{xapprox}
\eeq
where $c_n^{(approx)}$ are the approximate Fourier coefficients obtained with our method. As for the $\alpha_n$, also the 
$c_n^{(approx)}$ are analytical and take the form
\beq
c_n^{(approx)} = \sum_{m=0}^{N_{max}} \ \overline{c}_{nm} 
\eeq
where $\overline{c}_{nm}$ are the corrections of order $m$ to the Fourier coefficient corresponding to $\cos \left[(2 n+1) \ \Omega \ t\right]$.
The latter can be written as:
\beq
\overline{c}_{nm} &=&  \frac{\beta_{nm} \ A \ \left( A^2 \ \mu\right)^n}{\left(1 + \frac{3}{4}  \ A^2 \ \mu \right)^{n}} \ ,
\eeq
where $\beta_{nm}$ is a numerical coefficient. 
In Fig.~\ref{Fig00} we display the logarithm of $\beta$ corresponding to different Fourier modes and to different orders in our method: interestingly, 
we observe that also the $\beta$s decay exponentially with the order of the expansion.

\begin{table}
\caption{\label{table1} First few coefficients of the expansion of eq. (\ref{alpha}).}
\begin{center}
\begin{tabular}{|c|c|c|c|}
\hline
$n$ & $\kappa_n$ & $n$ & $\kappa_n$  \\
\hline
$2$ & $\frac{3}{128}$ & $12$ & $ \frac{19974549}{18014398509481984}$\\
$4$ & $\frac{51}{131072}$ & $14$ & $\frac{128255751}{2305843009213693952}$ \\
$6$ & $\frac{213}{16777216}$ & $16$ & $\frac{435036452211}{151115727451828646838272}$ \\
$8$ & $\frac{70515}{137438953472}$ & $18$ & 
$\frac{2950668677535}{19342813113834066795298816}$ \\
$10$ & $\frac{406179}{17592186044416}$ & $20$ &  $ \frac{163068192461619}{19807040628566084398385987584}$ \\
\hline
\end{tabular}
\end{center}
\end{table}

\begin{figure}
\includegraphics[width=9cm]{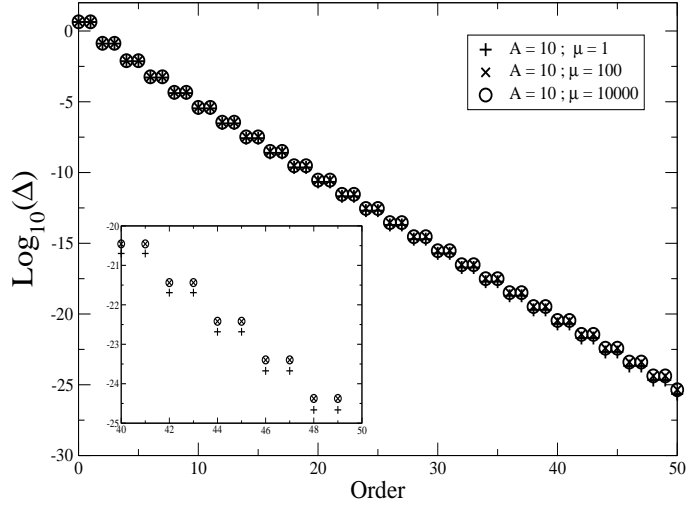}
\caption{Logarithm of the percentile error defined in eq. (\ref{eq:error}) for $A=10$ and different values of $\mu$ ($\mu >0$). }
\label{Fig1}
\end{figure}

We now present some  results obtained with our method. We will first consider the case of positive $\mu$.
In Fig.~\ref{Fig1} we plot the logarithm of the error defined the equation
\beq
\Delta \equiv \left| \frac{\Omega^2 - \Omega^2_{exact}}{ \Omega^2_{exact}} \right| \ \times 100 \ ,
\label{eq:error}
\eeq
where $\Omega^2$ is the squared frequency obtained with the LPLDE method and $\Omega^2_{exact}$ is the exact value.
The blow-up in the figure allows to better appreciate the small differences in the error corresponding to the 
different values of $\mu$ (we use $\mu = 10, 100,10000$). We notice that the error is practically unaffected by the size of $\mu$.
This result is clearly understood by eq. (\ref{alpha}), because the coefficients $\alpha_n$ clearly go to $0$ faster than the
$\kappa_n$ for $\mu >0$.

\begin{figure}
\includegraphics[width=16cm]{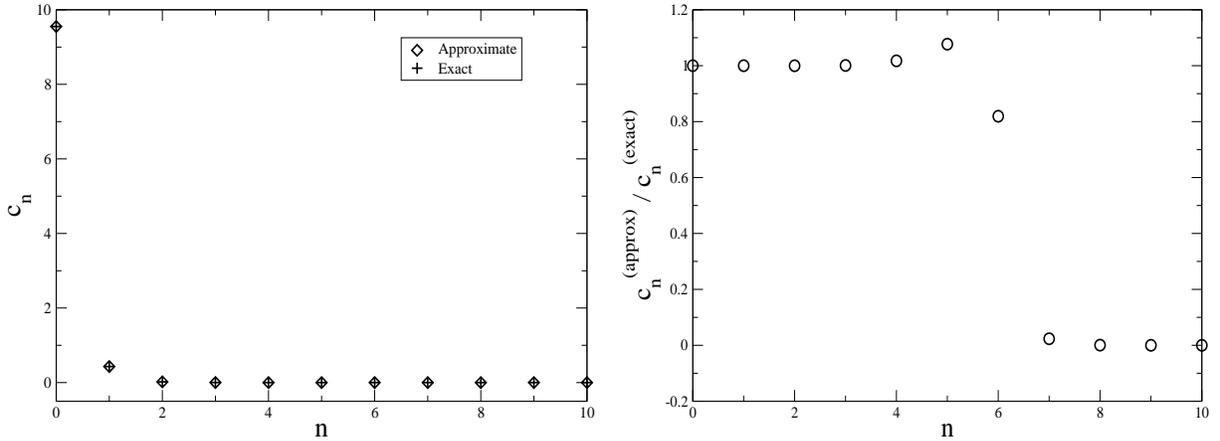}
\caption{Left plot: exact (numerical) and approximate coefficients of the Fourier series corresponding to 
$A= 10$ and $\mu = 10^4$; Right plot: ratio of the approximate to the exact coefficients.}
\label{Fig2}
\end{figure}

In  Fig.~\ref{Fig2} we compare the coefficients $c_n^{(approx)}$ of the approximate solution with the ones of the exact (numerical)
solution:
\beq
\label{fourier1}
x_{exact}(t) &=& \sum_{n=0}^\infty \ c_n^{(exact)} \ \cos \left[(2 n+1) \ \Omega \ t\right]   \ .
\eeq
Notice that the approximate series (\ref{xapprox}) is truncated at the maximum frequency $(2 N_{max}+1) \ \Omega_{max}$;
however this cutoff frequency can be increased by applying the method to higher orders\footnote{The results displayed in the Figure  
are calculated to order $50$.}.

Clearly our method is capable of reproducing the first few coefficients of the Fourier series with great accuracy. Although
the modes with higher frequency turn out to be poorly approximated, they don't affect  the overall quality of
the approximation, given the small size of their contributions.

We now consider $\mu = -1$. In this case the potential has maxima located at $x = \pm 1$ and 
an oscillatory behaviour is permitted only for amplitudes $A < 1$. $A = \pm 1$ are points of (unstable) equilibrium and therefore the period diverges 
in correspondence of these values.

\begin{figure}
\includegraphics[width=9cm]{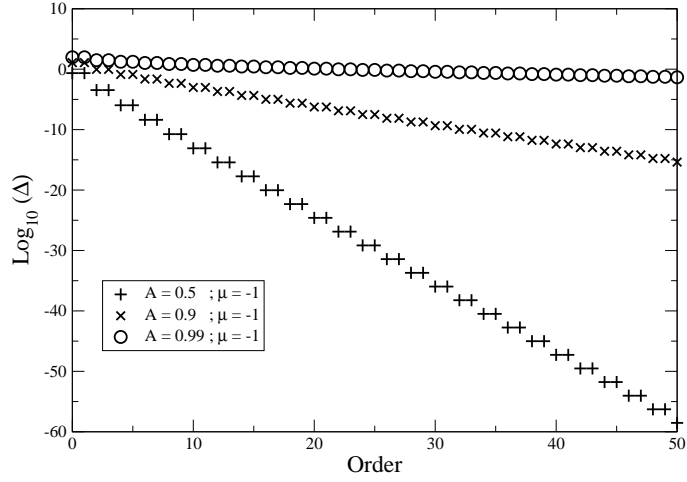}
\caption{Logarithm of the percentile error defined in eq. (\ref{eq:error}) for $\mu = -1$ and different amplitudes. }
\label{Fig3}
\end{figure}

By looking at Fig.~\ref{Fig3} we see that the slope of the logarithm of the error is now strongly dependent upon the amplitude, in contrast 
with the case previously analyzed. This result can be understood in view of eq.(\ref{alpha}): because $A^2 \ \mu$ is now negative, the sign
in the equation is changing order by order and the size of the denominator is smaller.

\begin{figure}
\includegraphics[width=16cm]{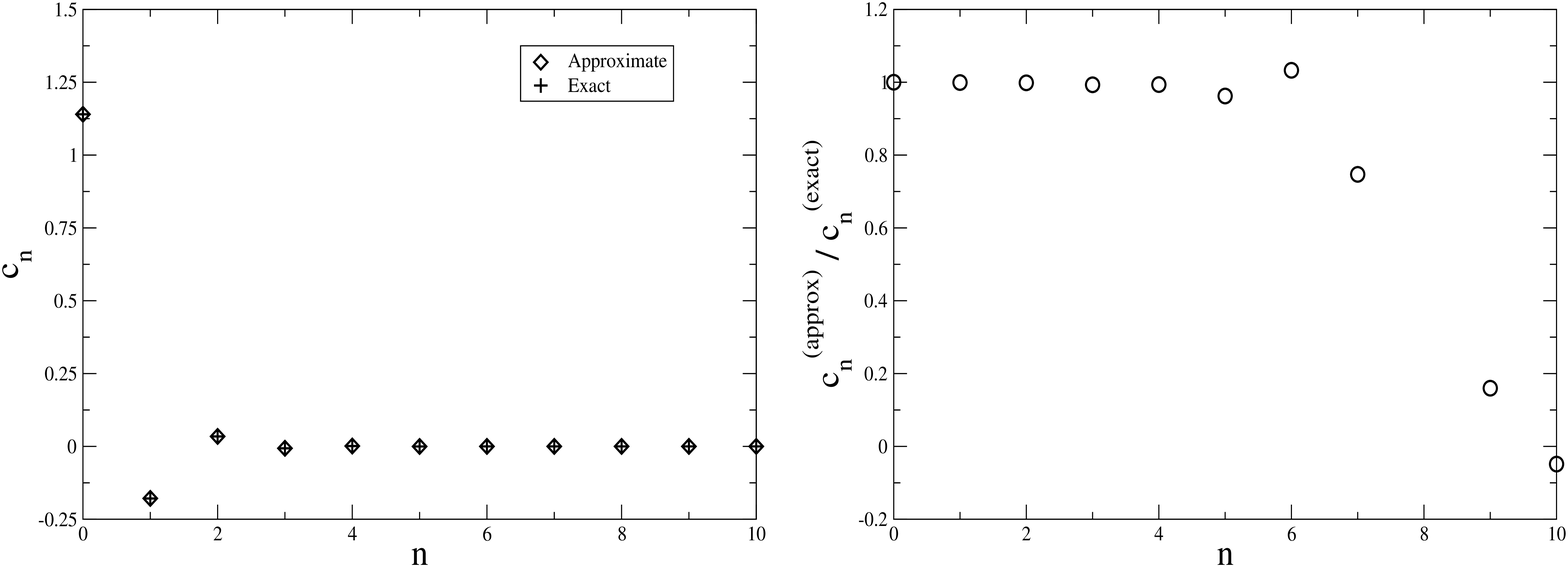}
\caption{Left plot: exact (numerical) and approximate coefficients of the Fourier series corresponding 
to $A= 0.99$ and $\mu = -1$; Right plot: ratio of the approximate to the exact coefficients.}
\label{Fig4}
\end{figure}

\begin{figure}
\includegraphics[width=16cm]{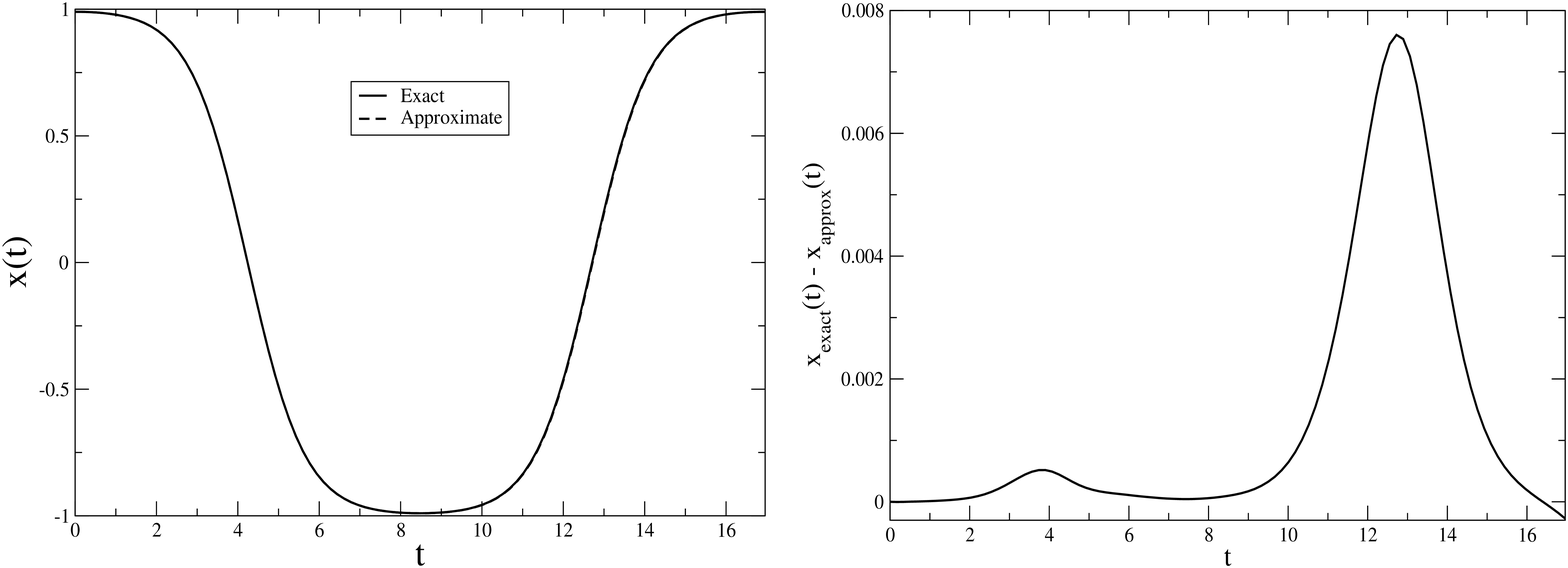}
\caption{Left plot: exact (numerical) and approximate solutions of the Duffing equation corresponding to 
$A= 0.99$ and $\mu = -1$; Right plot: the difference $x_{exact}(t)-x_{approx}(t)$.}
\label{Fig5}
\end{figure}

In Fig.~\ref{Fig4} and \ref{Fig5} we analyze the Fourier coefficients of the approximate and exact solutions and then plot the difference 
between the two as a function of time. We consider oscillations with an amplitude very close to the maximum, i.e. $A = 0.99$. Also in this case
we notice that the approximation works very well.

\section{Anharmonic potentials}
\label{sec:anharm}

The analysis carried out in the previous Section can be extended easily to more general anharmonic potentials of the form:
\beq
V(x) &=& \frac{1}{2} \ x^2 + \frac{\mu}{2 N} \ x^{2 N}
\eeq
with $N$ positive integer. In particular we study here the sextic and octic oscillators: we will see that many of the features
observed previously for the Duffing equation hold true also in this case. Remarkably the optimal value of $\lambda$ calculated to 
third order falls very close to the exact solution of the PMS condition to any given order. This result will allow
us to obtain fully {\sl analytical} expressions also in the case of sextic and octic oscillators.

We obtain indeed the expressions for the optimal values of $\lambda$:
\beq
\lambda_{sextic} = \sqrt{\frac{211 \ A^4 \ \mu}{312}} \ \ \ , \ \ \ \lambda_{octic} =  \sqrt{\frac{10885 \ A^6 \ \mu}{16896}}
\eeq

\begin{figure}
\includegraphics[width=16cm]{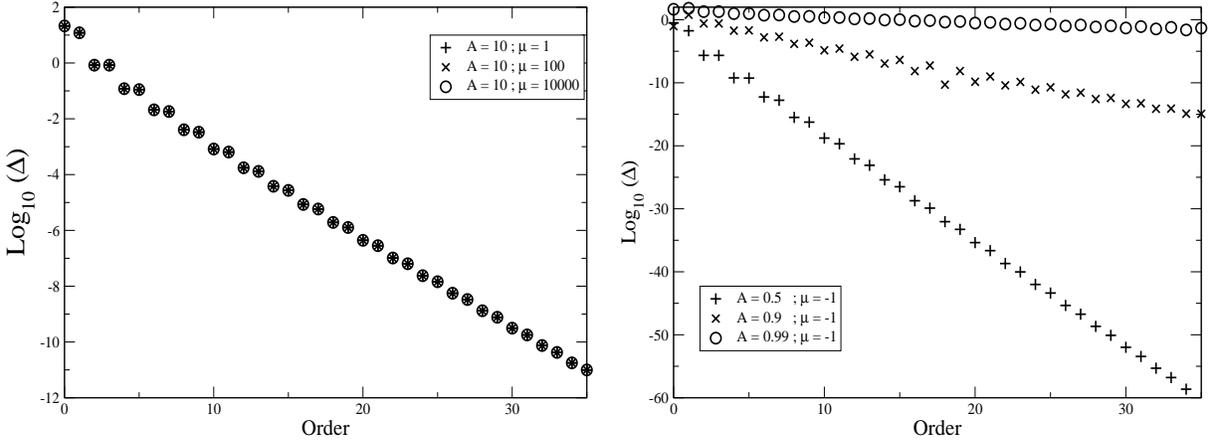}
\caption{Left plot: logarithm of the error over the squared frequency for the sextic oscillator with $\mu = 1, 100,10000$.
Right plot: logarithm of the error over the squared frequency for the sextic oscillator with $A= 0.5$,$0.9$ and $0.99$ and $\mu = -1$.}
\label{Fig6}
\end{figure}

\begin{figure}
\includegraphics[width=16cm]{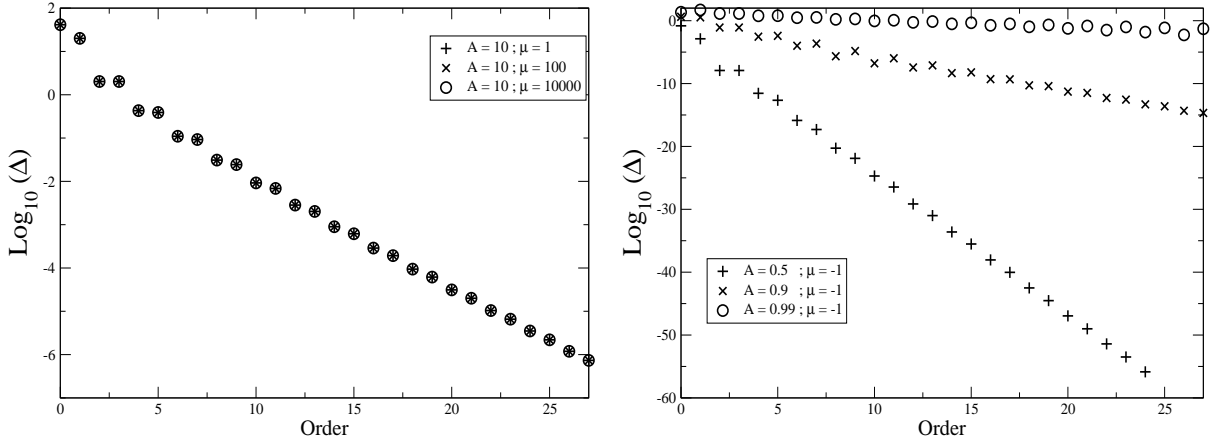}
\caption{Left plot: logarithm of the error over the squared frequency for the octic oscillator with $\mu = 1, 100,10000$.
Right plot: logarithm of the error over the squared frequency for the octic oscillator with $A= 0.5$,$0.9$ and $0.99$ and $\mu = -1$.}
\label{Fig7}
\end{figure}

By looking at Fig.~\ref{Fig6} and \ref{Fig7} we see that our method provides an excellent approximation both in the regime of positive and negatives
$\mu$: indeed we see that the error decays exponentially following a law which is very similar to the one observed in the Duffing oscillator.
These similarities reflect also in the behaviour of the Fourier coefficients, which however we will not display here.

\section{Van der Pol equation}
\label{sec:vdp}

We now come to consider the Van der Pol equation (see for example \cite{buo98}):
\beq
\frac{d^2x}{dt^2} + x = \mu \ \left( 1 - x^2 \right) \ \frac{dx}{dt}  \ .
\label{vdp}
\eeq

This equation possesses a limit cycle and can sustain periodic oscillations with definite period and amplitude depending upon the
constant $\mu$. However, unlike the cases treated before, eq. (\ref{vdp}) does not correspond to a conservative system and 
indeed the r.h.s of the equation either damps or enhances the oscillations depending upon the size of $x$. 

We will now apply our method  to this problem in the usual fashion by writing:
\beq
\Omega^2 \ \ddot{x} + \left(1+\lambda^2\right) \ x(\tau) = \delta \left[ \mu \ \Omega \ \left( 1 - x^2 \right) \ \dot{x} + 
\lambda^2 \ x(\tau) \right] \ .
\label{vdp1}
\eeq
where the expansion 
\beq
\Omega = \sum_{n=0}^{N_{max}} \ \delta^n \ \gamma_n \ \ \ , \ \ \ x(t) = \sum_{n=0}^{N_{max}}  \ \delta^n \ x_n(\tau)  \ .
\eeq
is assumed. The coefficients $\gamma_n$ are fixed by the requirement that the resonant terms at a given order $n$ vanish.

\begin{figure}
\includegraphics[width=16cm]{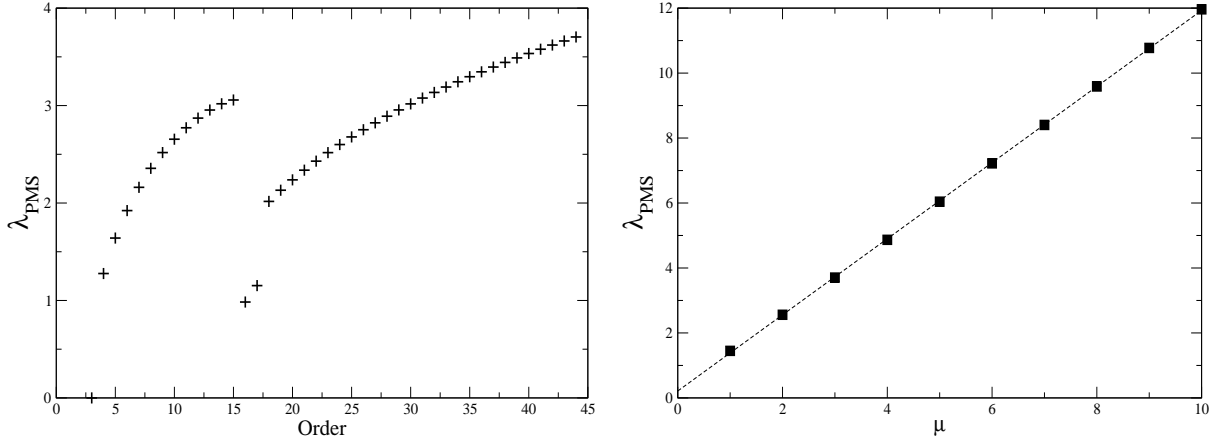}
\caption{Left plot:Optimal value of $\lambda$ as a function of the order for $\mu = 3$. Right plot: Optimal value of $\lambda$ to order $44$ as
a function of $\mu$. The dashed line is the fit $\lambda  = 0.212599 + 1.17166 \ \mu$. }
\label{Fig8}
\end{figure}

As we can see from the left plot of Fig.~\ref{Fig8} the optimal value of $\lambda$ turns out to be strongly order dependent 
for a given $\mu$  unlike in the cases considered previously. 
However, at a fixed order, we observe that the optimal value of $\lambda$ grows linearly with $\mu$ (right plot). The dashed line in the right plot
is the fit $\lambda   = 0.212599 + 1.17166 \ \mu$. This behaviour of $\lambda$ allows to obtain a semi-analytical formula both for the 
period and for the solution which works for moderate values of $\mu$, where the LP method is already not applicable.
 
In Table~\ref{table2} we report the period of the solution of the Van der Pol equation as obtained in our method (to order $44$) 
for $\mu$ ranging from $1$ to $10$ and we compare it with the numerical results of \cite{stra73}. 
Notice that, at the largest value of $\mu$ considered here ($\mu = 10$), we obtain an error of $12 \ \%$. 
Much smaller errors are obtained at lower values of $\mu$. 

\begin{figure}
\includegraphics[width=9cm]{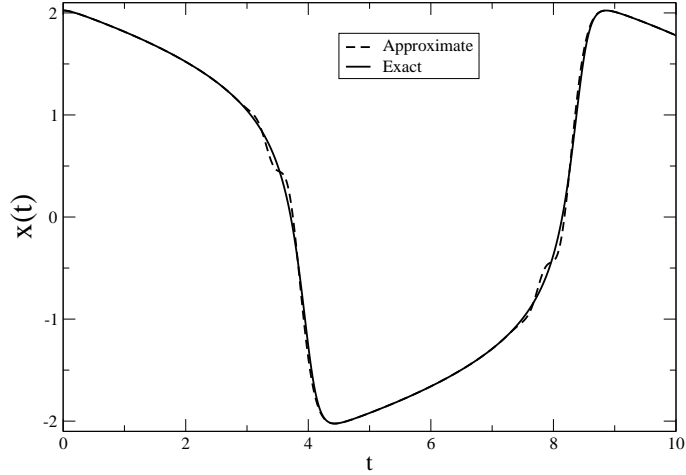}
\caption{Approximate and exact solutions of the Van der Pol equation to order $43$ for $\mu = 3$. }
\label{Fig10}
\end{figure}

In Fig.~\ref{Fig10} we display the approximate (to order $43$) and exact (numerical) solutions of the Van der Pol equation for $\mu = 3$.
For such a large value of $\mu$ our method provides an excellent approximation both to the period and to the function. Notice that
in this regime the LP method is not applicable.

\begin{table}
\caption{\label{table2} Comparison between the approximate and exact period of the Van der Pol equation. }
\begin{center}
\begin{tabular}{|c|c|c|c|c|c|}
\hline
$\mu$ & $T_{approx}$ & $T_{exact}$ & $\mu$ & $T_{approx}$ & $T_{exact}$ \\
\hline
$1$ & $6.66328685$ & $6.66328686$ & $6$ & $12.83944539$   & $13.06187474$ \\
$2$ & $7.62995604$  & $7.62987448$ & $7$ & $14.03245644$   & $14.53974774$ \\
$3$ & $8.86085271$   & $8.85909550$ & $8$ & $15.09907603$   & $16.03817623$ \\
$4$ & $10.19701347$  & $10.20352369$ & $9$ & $16.02685408$    & $17.55218414$ \\
$5$ & $11.54602911$   & $11.61223067$ & $10$ & $16.8128186$   & $19.07836957$ \\
\hline
\end{tabular}
\end{center}
\end{table}

\begin{figure}
\includegraphics[width=16cm]{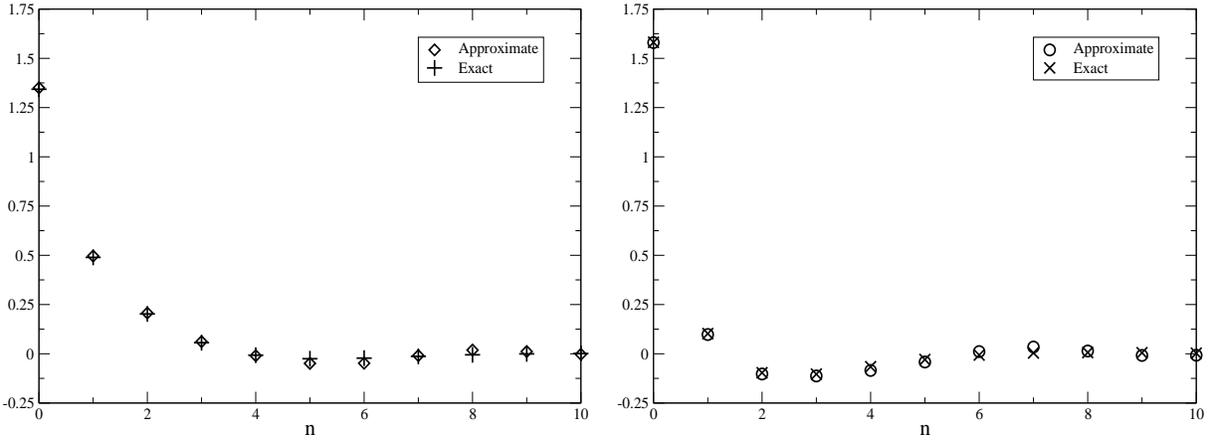}
\caption{Approximate and exact Fourier coefficients of the Van der Pol equation calculated at order $43$ for $\mu = 3$. Left plot: coefficients 
of the $\cos$;  Right plot: coefficients of the $\sin$.}
\label{Fig11}
\end{figure}

\section{Conclusions}
\label{sec:concl}

In this work we have extended the analysis carried out in \cite{aapla03} and we have applied the LPLDE method 
to much higher orders: we have observed that the results obtained converge quickly to the exact values, with  errors which 
decay exponentially with the perturbative order. This behaviour was observed for the Duffing equation, for the sextic and
octic oscillator. Remarkably the results obtained to {\sl any given order} with our method are fully analytical. 

The method was then tested on the Van der Pol equation where the calculations were performed up to order $44$. In this case, 
we were able to obtain good results for moderate values of $\mu$ ($\mu \approx 3$), where the LP method already fails. 
These results suggest that the extension of the method to higher orders could extend the region of application to larger values 
of $\mu$. Moreover, although the optimal value of the variational parameter is not known analytically, as in the previous case, 
the use of a $\mu$ dependent fit allowed us to obtain also in this case a semi-analytical approximation for the frequency 
(period) which holds in the region of moderate $\mu$. 

It remains to explore the application of the present method to a wider class of nonlinear problems with
periodic solutions, as well as studying the Van der Pol equation to higher orders. This will be done in future work.

\section{Acknowledgments}

The Authors would like to thank A.Aranda and R. S\'aenz for useful discussions.
They also  acknowledge the support of the ``Fondo Alvarez-Buylla'' of the University of Colima and of Conacyt grant no. C01-40633/A-1.


\newpage


\end{document}